\documentclass[referee]{aa}
\usepackage{graphics}

\newcommand{\bea}{\begin{eqnarray}}
\newcommand{\eea}{\end{eqnarray}}
\newcommand{\be}{\begin{equation}}
\newcommand{\ee}{\end{equation}}
\begin{document}
\thesaurus{06    
        (13.07.3; 13.07.2; 13.18.3; 09.19.2; 02.02.1)}
\title{An inquiry into the nature of the gamma-ray source 3EG J1828+0142}
\author{Brian Punsly \inst{1}, Gustavo E. Romero \inst{2,}\thanks{Member
of CONICET}, Diego F. Torres \inst{2}, J.A. Combi \inst{2, \star}
} \offprints{G.E. Romero}

\institute{4014 Emerald Street, No. 116, Torrance, CA 90503, USA
\and Instituto
 Argentino de Radioastronom\'{\i}a, C.C.5,
(1894) Villa Elisa, Bs. As., Argentina }

\authorrunning{Punsly et al.}
\titlerunning{The gamma-ray source 3EG J1828+0142 }

\date{\today}

\maketitle


\begin{abstract}
The unidentified, low-latitude, $\gamma$-ray source 3EG J1828+0142
presents high levels of variability and a steep spectral index
$\Gamma\sim2.7$. Here we propose a model for this source where the
high-energy emission is produced by a galactic Kerr-Newman black
hole. The model takes into account electron-positron annihilation
losses in the calculation of the expected spectral energy
distribution and can reproduce the observational features,
including the absence of a strong radio counterpart. We also
report the discovery of a nearby supernova remnant that could be
associated with the original supernova explosion that created the
black hole. Several faint radio sources were also detected in the
radio field within the inner $\gamma$-ray confidence contour and
their spectral index estimated. Some of these sources could be the
expected weak radio counterpart. \keywords{Gamma rays: theory --
Gamma rays: observations -- Radio continuum: ISM -- ISM: supernova
remnants -- Black hole physics}

\end{abstract}

\section{Introduction}

During its lifetime, the EGRET instrument on board the Compton
Gamma Ray Observatory detected 170 gamma-ray point sources which
are not clearly identified with known objects at lower frequencies
(Hartman et al. 1999). About a half of these sources are
concentrated near the galactic plane, suggesting that they have a
relatively local origin (Romero et al. 1999a, Gehrels et al.
2000).

Different kinds of possible counterparts have been suggested for
the galactic population of gamma-ray sources: supernova remnants
(SNRs) in interaction with molecular or atomic clouds (e.g.
Sturner \& Dermer 1995, Esposito et al. 1996, Combi et al. 1998),
massive stars with strong stellar winds (e.g. Romero et al.
1999a), young pulsars (e.g. Yadigaroglu \& Romani 1997, Zhang et
al. 2000), and isolated black holes (e.g. Dermer 1997, Punsly
1998a,b).

Recent variability analysis of the data in the Third EGRET catalog
by Torres et al. (2000) show that many of the sources at low
galactic latitudes display high levels of variability, confirming
the results found by McLaughlin et al. (1996) for the sources of
the Second EGRET catalog. The conjunction of steep gamma-ray
spectral indices and strong variability in a few sources seems to
suggest the existence of an entirely new population of high-energy
sources in the Galaxy. One of such sources is 3EG J1828+0142,
located at $(l,\;b)\approx(31.9\degr,\;5.78\degr)$. Hartman et al.
(1999) suggest that it could be an Active Galactic Nucleus (AGN),
although there is no strong radio blazar within the 95 \%
confidence contour. The source presents a steep spectral index
with a value $\Gamma=2.76\pm0.39$ and variable gamma emission. If
we introduce the variability index $I=\mu_{\rm s}/<\mu>_{\rm p}$,
where $\mu_{\rm s}=\sigma/<F>$ is the fluctuation index of the
gamma-ray source and $<\mu>_{\rm p}$ is the averaged fluctuation
index of all known gamma-ray pulsars (which are usually considered
as a non-variable population), we found that $I=5.5$, a clear
indication of strong variability.

In this paper we propose that this source could be a magnetized
black hole originated in a relatively recent supernova explosion.
We have used background filtering techniques to isolate the radio
image of an extended SNR overlapped to the gamma-ray source from
the contaminating diffuse emission of the Galaxy in radio
observations at two frequencies. Both the variability and the
steep spectral index of the gamma-ray radiation argue against the
possibility that the compact object left after the explosion be a
pulsar. We present a model for a magnetized black hole that can
reproduce the observed high-energy spectrum and predicts an
intense electron-positron annihilation line that could be detected
in a couple of years by the SPI spectrometer of the INTEGRAL
satellite.

In the next section we describe the data analysis technique. Then,
we present our results, briefly discuss the probability of a
chance association of the SNR and the gamma ray source, and
outline the black hole model. Finally, we discuss the predictions
that can test our proposal.

\section{Data analysis and results}

\begin{figure*}
\resizebox{7cm}{!}{\includegraphics{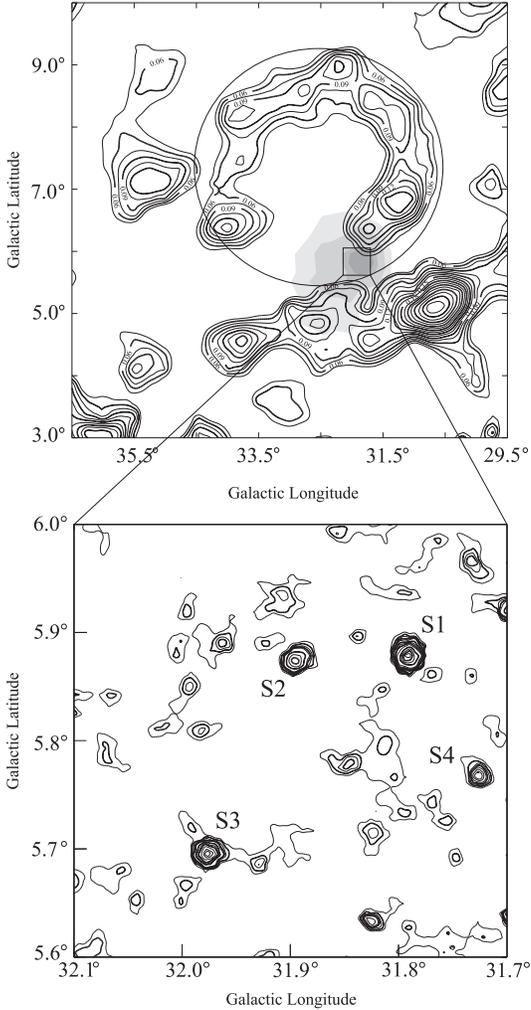}}\hfill
\parbox[b]{55mm}{ \caption{{\bf Upper panel:} Filtered large-scale
radio emission around the $\gamma$-ray source 3EG J1828+0142 at
1.4 GHz. Radio contours are shown in steps of $0.015$ K, starting
from $0.045$ K (noise level is $0.02$ K). The $\gamma$-ray
probability contours are superposed in gray scale. {\bf Lower
panel:} Small-scale 1.42-GHz emission obtained from the VLA sky
survey for the region within the inner probability contour of the
$\gamma$-ray source. Radio contours are in steps of 1 mJy,
starting from 1 mJy. The characteristics of the four point sources
labeled from S1 to S4 are detailed in Table 2.} \label{fig1}}
\end{figure*}

We have studied the surroundings of  3EG J1828+0142 using radio
data from the surveys by Haslam et al. (1981) and Reich \& Reich
(1986). The background filtering method developed by Sofue \&
Reich (1979) was applied to remove the diffuse galactic emission
that hides weak and extended radio structures at low galactic
latitudes. The procedure and its application to find low surface
brightness SNRs is described in detail by Combi et al. (1998,
1999). In the present case we have applied a gaussian filtering
beam of $90'\times90'$ to 0.408- and 1.42-GHz maps of a $7\degr
\times 7\degr$ field around  3EG J1828+0142, finding out a
previously unnoticed shell-type structure centered at
$(l,\;b)\approx(32.6\degr,\;7.3\degr)$. It is a large ($4\degr
\times 4\degr$) source with a total flux density at 1.42 GHz of
$18.2\pm2.1$ Jy and a nonthermal spectral index
$\alpha=0.72\pm0.18\;(S(\nu)\propto\nu^{-\alpha})$, very similar
to other shell-type SNRs (see Green 1998). In Figure \ref{fig1},
upper panel, we show the filtered 1.42-GHz image of the source.
The large continuum sources at the South of the new remnant
candidate are probably related to the North Polar Spur (see
Jonas's 1999 large-scale map of the region at 2.326 GHz). Table 1
lists the main characteristics of the new SNR candidate. The
distance of $\sim 940$ pc has been estimated using the $\Sigma -D$
relationship derived by Allakhverdiyev et al. (1988) for low
surface brightness SNRs, and it should be considered just as a
very rough estimate. In any case, the large apparent size of the
remnant suggests that it is nearby.

If we assume a standard energy release of $ \sim 0.4 \times
10^{51}$ erg for the supernova explosion and a typical intercloud
density of $0.1$ cm$^{-3}$ for the interstellar medium (Spitzer
1998), we get from the Sedov solutions that the age of the remnant
is $\sim 43\,500$ yr. A transverse velocity of $\sim 730$ km
s$^{-1}$ is then required for the compact object left by the
explosion in order to be currently located near the outer boundary
of the SNR. A denser ambient medium would imply, of course,
smaller velocities. For instance, with a density $\sim 1$
cm$^{-3}$ we get a more reasonable velocity of $\sim 240$ km
s$^{-1}$. It should be also taken into account that the
determination of the distance has an uncertainty of $\sim 50$ \%,
so even lower velocities are possible. Lyne \& Lorimer (1994) have
recently estimated that the space velocity of pulsars at birth has
a mean value of $450\pm 90$ km s$^{-1}$, which is a factor three
higher than earlier estimates (e.g. Lyne, Anderson \& Salter
1982). Stellar black holes should present a velocity
distribution similar to that of pulsars.

\begin{table}
\caption[]{Measured properties of the SNR}
\begin{tabular}{ l c }
 \cr \hline
  Property    &  Value   \\
\hline Galactic coordinates (center) &
($+32.6^{\circ}$,$7.3^{\circ}$) \\ Angular size (deg $\times$deg)
& $4.0 \times 4.0$ \\ Flux density (408 MHz) & $44.7 \pm 6.5$ Jy
\\ Flux density (1420 MHz) & $18.2 \pm 2.1$ Jy  \\ Average
spectral index  & $0.72 \pm 0.18$ \\ Distance (pc)  & $\sim 940$
\\ \hline
\end{tabular}
\end{table}


In order to explore the small-scale radio emission within the
inner probability contours of the gamma-ray source we have used
the NVSS Sky Survey with an angular resolution of 45 arcseconds
(Condon et al. 1998). There are just four point sources with
fluxes above 10 mJy within the 65 \% probability contour of 3EG
J1828+0142. They have been labeled from S1 to S4 in Figure
\ref{fig1}, lower panel. All these sources are nonthermal, with
flux densities comprised between $\sim 5$ and 100 mJy at 1.4 GHz.
Sources S1, S2 and S4 do not present resolved structure, whereas
the source S3 has two weak extensions towards Northeast and
Southwest. It is hard to say, however, whether these features are
artifacts due to confusion with weak, neighboring sources, or real
components of S3. The characteristics of the four sources are
summarized in Table 2. The sources S2, S3 and S4 have no entry in
any point source catalog at present. We have estimated lower
limits to their radio spectral index using the 5 GHz survey by
Condon et al. (1994), which is sensitive down to 5 mJy.

\begin{table*}
\caption[]{Characteristics of the point radio sources inside the
$\gamma$-ray contour}
\begin{tabular}{ c c c c c c }
\cr \hline Source & ($l$, $b$) & $F_{\rm 1.4
GHz}$ & $F_{\rm 365 MHz}$ & $\alpha$  & ID  \\
       &  (deg, deg) &  (mJy) & (mJy) &  &  \\
\hline S1 & (31.79,$+5.87$) & 78.4  & 275.4 & $0.9$ & TXS
1825+016$^1$
\\ S2 & (31.89,$+5.88$) & 13.8  & -     & $>0.8$    &     -       \\ S3
& (31.98,$+5.69$) & 25.1  & -    & $>1.2$      & - \\ S4 &
(31.73,$+5.77$) & 7.3 & - &$>0.3$ &- \\ \hline \multicolumn{6}{l}
{$^1$ Douglas et al. (1996)}
\end{tabular}
\end{table*}

\section{Probability of chance association}

In order to make quantitative estimates of the probability of a
pure chance superposition between the SNR and the gamma-ray source
we have adopted the numerical code developed by Romero et al.
(1999b) and used by Romero et al. (1999a) to study the positional
association of unidentified EGRET sources with various populations
of galactic objects. The code calculates angular distances between
different kinds of celestial objects contained in selected
catalogues, and establishes the level of positional correlation
between them. Numerical simulations using large numbers of
synthetic populations are then performed in order to determine the
probabilities of pure chance associations. When generating
synthetic populations of $\gamma$-ray sources the distribution in
galactic latitude is constrained to be the same as the one
actually observed for the 3EG sources. This is necessary in order
to obtain reliable results since the distribution of the 3EG
sources is non-isotropic, with a strong concentration towards the
galactic plane. The reader is referred to the paper by Romero et
al. (1999a) for further details of the simulation code.

Our results for the case of 3EG J1828+0142 indicates that the
probability of finding by chance a gamma-ray source with its 95\%
confidence contour within the outer boundary of the SNR is $7.0
\times 10^{-2}$ . The probability of finding a variable gamma-ray
source (estimated taken into account the actual fraction of
variable sources in the 3EG catalog according to Torres et al.
2000) is lower: $1.0 \times 10^{-2}$. These values are not too
compelling, but if we calculate the probability of the gamma-ray
source being a background AGN seen through the galactic plane and
associated by chance with the SNR, we get a value of $\sim 4\times
10^{-6}$, which is significantly lower. In making this calculation
we have extrapolated the isotropic population of already detected
gamma-ray emitting AGNs towards the region obscured by the
galactic disk emission, performing simulations with no gamma-ray
source density gradient towards the plane.

\section{Magnetized black hole model }

Kerr-Newman isolated black holes are interesting candidates to
produce variable gamma-ray sources in the Galaxy (Punsly 1998a,
1998b). The configuration of a simple axisymmetric magnetosphere
around a maximally rotating black hole attains a minimum energy
configuration when the hole and the magnetosphere have equal and
opposite charge. Punsly (1998a) has shown that the magnetospheric
charge can be supported in a stationary orbiting ring or disk. The
entire magnetized system is stable only in an isolated
environment, otherwise accretion onto the hole would disrupt the
ring and its fields. Kerr-Newman black holes are charged similarly
to neutron stars in pulsars. But unlike neutron stars, black holes
have no solid surface and consequently no thermal X-ray emission
is expected. These objects can support strong magnetized bipolar
winds in the form of jets where gamma-ray emission is originated
by the inverse Compton mechanism (Punsly 1998a). Since both magnetic
and rotation axes are always aligned in them, their emission is
nonpulsating (NP) and for such a reason they have been called NP
black holes (Punsly 1999).

We propose that 3EG J1828+0142 could be a NP black hole created by
the same supernova explosion that produced the nearby SNR. In what
follows we present a specific model that reproduces the steep
gamma-ray spectrum observed at EGRET energies. The model takes
into account electron-positron annihilations in the inner jet,
self-Compton cooling of the relativistic leptons, and synchrotron
emission of the outer jet, in such a way that it provides concrete
predictions for different wavebands that can be tested in the near
future.

We shall follow the treatment given by Punsly (1998a), considering
a black hole mass $M=4$ $M_{\sun}$ and a polar magnetic field
strength $B=10^{11}$ G. The inner jet begins at the inner light
cylinder (located at a cylindrical radius $r_{\rm lc}$ from the
symmetry axis):
\begin{equation}
r_0=r_{\rm lc}=8\times10^7\;\;\;{\rm cm};
\end{equation}
this is at a distance $R_0$ from the event horizon:
\begin{equation}
R_0=10^8\;\;\;{\rm cm}.
\end{equation}

The radius of the jet $r$ is given in terms of the axial
displacement from the black hole, $R$, as:
\begin{equation}
r=(R/R_0)^{\epsilon}\;r_0,\;\;\;\;\epsilon=0.2.
\end{equation}
Thus, the inner jet is tightly collimated. Its length is given by:
\begin{equation}
R_{\rm max}=1.5\times10^{12}\;\;\;{\rm cm}.
\end{equation}
The Doppler enhancement factor is constant and assumed to be:
\begin{equation}
\delta=2.5.
\end{equation}
The magnetic field and the particle number density vary with $R$
as:
\begin{equation}
B=100\;(R/R_0)^{-\epsilon} \;\;\;{\rm G}
\end{equation}
and
\begin{equation}
N=N_{\Gamma}\;(R/R_0)^{-\epsilon},
\end{equation}
where
\begin{eqnarray}
N_{\Gamma}&=&7.5\times10^{13}\;\;\;\;{\rm cm^{-3}},\\
n&=&\int^{\gamma_{\rm max}}_{\gamma_{\rm min}} N_{\Gamma}
\gamma_{\rm th}^{-\Gamma} \;d\gamma_{\rm th\;}, \\ \Gamma&=&2.
\end{eqnarray}
The maximum thermal Lorentz factor, $\gamma_{\rm th}=\gamma_{\rm
max}$, is:
\begin{equation}
\gamma_{\rm max}=1.5\times10^4\;(R/R_0)^{-0.225}.
\end{equation}

Gamma-rays are produced in this inner jet by electron-positron
annihilations and self-Compton emission. The annihilation
luminosity is enhanced by Doppler boosting in the jet as:
\begin{equation}
L^{\rm ann}_{\rm obs}= \delta^3 L_0^{\rm ann},
\end{equation}
where
\begin{eqnarray}
 L_0^{\rm ann}&=&(3/32) \sigma_T
c (\Gamma-1)^2 n^2 m_e c^2 V\times \nonumber\\ &&
\left[\frac{2}{(\Gamma-1/2)(\Gamma+1/2)}+\frac{2(2\Gamma-1)}{\Gamma^2
(\Gamma-1)^2 }\right] \label{ann}
\end{eqnarray}
(see Roland \& Hermsen 1995, noticing the wrong exponent in their
Eq. 4, where it should be no dependency on $\alpha$). In the above
Eq. (\ref{ann}), $n$ is the number density of electron-positron
pairs, $V$ is the volume where the annihilations occur, and
$\sigma_T$ is the Thomson cross section. The peak of the
annihilation line will be at an energy:
\begin{equation}
E_{\rm max}=\delta \gamma_{\rm min} 0.511\; {\rm MeV},
\end{equation}
which in our model corresponds to 6.4 MeV, assuming $\gamma_{\rm
min}\approx 5$. The spectral shape of the annihilation line is as
in B\"ottcher \& Schlickeiser (1996). Notice, however, that these
authors consider an external source of photons for the inverse
Compton radiation and, consequently, their conclusions on the
minimum leptonic number density in the jet do not apply to our
case.

\begin{figure}
\resizebox{8.5cm}{!}{\includegraphics{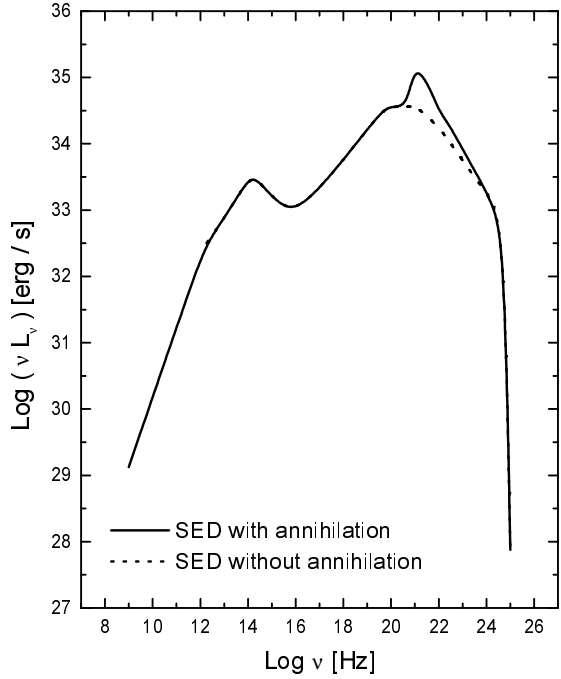}} \caption{Spectral
energy distribution for the magnetized black hole model.}
\label{fig2}
\end{figure}

The outer jet, which is responsible for the bulk of the
synchrotron emission, can be parameterized as a direct
extrapolation of the inner jet:

\begin{equation}
R^{\rm out}_0=1.5 \times 10^{12}\;\;\;\;{\rm cm},
\end{equation}

\begin{equation}
r^{\rm out}_0=5.5 \times 10^{8}\;\;\;\;{\rm cm},
\end{equation}

\begin{equation}
r^{\rm out}=(R/R^{\rm out}_0)^{1.5}\;r^{\rm out}_0,
\end{equation}

\begin{equation}
R^{\rm out}_{\rm max}=1.5\times10^{15}\;\;\;{\rm cm},
\end{equation}

\begin{equation}
\Gamma=2, \;\; \alpha=(\Gamma-1)/2=0.5,
\end{equation}

\begin{equation}
\delta=2.5 (R/R^{\rm out}_0)^{-0.1}, \label{delta}
\end{equation}

\begin{equation}
B=14.6\;(R/R^{\rm out}_0)^{-1.5} \;\;\;{\rm G},
\end{equation}

\begin{equation}
N_{\gamma}=1.1\times10^{13}(R/R^{\rm out }_0)^{-3.0}\;\;\;\;{\rm
cm^{-3}},
\end{equation}

\begin{equation}
\gamma_{\rm max}=1.72 \times 10^3(R/R^{\rm out}_0)^{-0.25}.
\end{equation}

The bulk of the gamma-ray emission is originated in the inner jet.
We have computed its emissivity using the synchrotron self-Compton
(SSC) formalism developed for AGNs by Ghisellini, Maraschi \&
Treves (1985) and adapted to magnetized black holes by Punsly
(1998b). The computed spectral energy distribution (SED) is shown
in Figure (\ref{fig2}). Notice that the radio luminosity is low,
so no strong point-like counterpart is expected at centimeter
wavelengths. The radio jets and the terminal radio lobes should
appear as a weak ($<30$ mJy at 5 GHz) source with an angular size
of a few arcseconds. Any of the sources in Table 2, then, are
potential counterparts.

At a few MeV, the gamma-ray annihilation luminosity exceeds the
SSC emission and the spectrum presents a broad peak. The pair
annihilation contribution produces a steepening in the spectrum,
which presents an index $\Gamma\sim2.7$ in the EGRET energy band,
consistent with the observations.

\section{Discussion and conclusions}

We have shown in the previous section that when electron-positron
annihilation effects are taken into account in the bipolar
magnetically dominated wind ejected by a Kerr-Newman black hole in
isolation, a steep-spectrum gamma-ray source with no strong radio
counterpart can be produced. The wind, which is tightly collimated
into a jet, possibly experiences strong dissipation near the outer
light cylinder (i.e. at $\sim 8\times10^7$ cm from the black hole)
due to plasma instabilities and shocks. These dissipative
processes can load the jet with hot plasma. If the inner jet were
to propagate with a bulk Lorentz factor of 5 and were viewed at 20
degrees off the axis, a Doppler factor of 2.5 would be attained as
in Eq. (\ref{delta}). The jet is likely to wiggle either as a
consequence of interactions with the enveloping medium or by
firehose instabilities. If the jet were to wobble $\pm$ 5 degrees
from its nominal value of $20\degr$, then the gamma-ray luminosity
would vary by a factor of 9. Thus, extreme variability is expected
in our model.

There are two main types of observations that could confirm in the
immediate future our proposal on the nature of the source 3EG
J1828+0142. On the one hand, VLBA observations of the sources in
Table 2 could reveal some structure unresolved in the VLA images
and show evidence of the twin jets and their end points in the
ISM, if there is a NP black hole producing 3EG J1828+0142. On the
other hand, observations with the gamma-ray spectrometer SPI in
the forthcoming INTEGRAL mission should clearly show evidence of
the electron-positron annihilation line, which has an expected
luminosity of $\sim 1.8 \times 10^{35}$ erg s$^{-1}$.

X-ray observations with Chandra observatory also could be very
useful to identify the lower frequency counterpart of 3EG
J1828+0142. The expected luminosity of the NP black hole in the
Chandra energy range is $\sim10^{34}$ erg s$^{-1}$, with a
spectral index $\Gamma=1.5\pm0.25$. It should appear as a point
source positionally coincident with one of the nonthermal radio
sources detected in the field.

The fact that the point-like radio sources found within the inner
confidence contour of the gamma-ray source have, in most cases,
steeper spectral
index than the canonical value of $0.5$ expected from the outer
jet can be explain by the contribution from the two lobes of radio
emission formed at the points where the outer jets end,
approximately at 0.1 pc from the black hole. The emission from
these lobes, which is not taken into account in the SED shown in
Figure \ref{fig2}, has been modeled by Punsly (1998b). The radio
spectrum from these regions is expected to be steeper than the
emission from the jet, with values of $\alpha\sim1.15$ at 1 GHz
(Punsly 1998b).

The recent variability analysis of the unidentified gamma-ray
sources in the Third EGRET catalog carried out by Torres et al.
(2000) shows that the most variable sources near the galactic
plane tend to present steep indices $\Gamma>2.5$. The model
presented here is capable of explaining the association of such
indices with high levels of variability and absence of strong
radio counterparts in a galactic source. The discovery of a nearby
SNR, also reported in this work, provides additional support to
the idea of a young black hole behind the source detected by
EGRET.

\begin{acknowledgements}

 This work has been supported by the Argentine agencies CONICET (PIP 0430/98)
 and ANPCT (PICT 98 No. 03-04881), as well as by Fundaci\'on Antorchas
(through funds granted to GER and DFT).

\end{acknowledgements}


{}

\newpage

\end{document}